\documentclass[lettersize,journal]{IEEEtran}
\usepackage{amsmath,amsfonts}
\usepackage{algorithmic}
\usepackage{algorithm}
\usepackage{array}
\usepackage[caption=false,font=normalsize,labelfont=sf,textfont=sf]{subfig}
\usepackage{textcomp}
\usepackage{stfloats}
\usepackage{url}
\usepackage{verbatim}
\usepackage{graphicx}
\usepackage{cite}
\usepackage{amsfonts}       
\usepackage{nicefrac}       
\usepackage{microtype}      
\usepackage{xcolor}         
\usepackage{amsmath}
\usepackage{multicol}
\usepackage{multirow}
\usepackage{adjustbox}
\usepackage{diagbox}
\usepackage{caption}
\usepackage{cuted}
\usepackage{longtable}
\usepackage{booktabs}
\usepackage[numbers]{natbib}
\usepackage{graphicx}
\usepackage{subcaption}

\newcommand{\AlgComment}[1]{\hfill$\triangleright$ \textit{#1}}
\begin{document}

\title{Fed-AugMix: Balancing Privacy and Utility\\ via Data Augmentation}

\author{Haoyang Li, Wei Chen, and Xiaojin Zhang
\thanks{Haoyang Li, and Xiaojin Zhang are with the National Engineering Research Center for Big Data Technology and System, Services Computing Technology and System Laboratory, Cluster and Grid Computing Laboratory, School of Computer Science and Technology, Huazhong University of Science and Technology, Wuhan 430074, China (e-mail: u202115372@hust.edu.cn; xiaojinzhang@hust.edu.cn)}
\thanks{Wei Chen is with the the School of Software Engineering, Huazhong University of Science and Technology, Wuhan 430074, China (e-mail: lemuria\_chen@hust.edu.cn)}
}



\maketitle

\begin{abstract}
Gradient leakage attacks pose a significant threat to the privacy guarantees of federated learning. While distortion-based protection mechanisms are commonly employed to mitigate this issue, they often lead to notable performance degradation. Existing methods struggle to preserve model performance while ensuring privacy. To address this challenge, we propose a novel data augmentation-based framework designed to achieve a favorable privacy-utility trade-off, with the potential to enhance model performance in certain cases. Our framework incorporates the AugMix algorithm at the client level, enabling data augmentation with controllable severity. By integrating the Jensen-Shannon divergence into the loss function, we embed the distortion introduced by AugMix into the model gradients, effectively safeguarding privacy against deep leakage attacks. Moreover, the JS divergence promotes model consistency across different augmentations of the same image, enhancing both robustness and performance. Extensive experiments on benchmark datasets demonstrate the effectiveness and stability of our method in protecting privacy. Furthermore, our approach maintains, and in some cases improves, model performance, showcasing its ability to achieve a robust privacy-utility trade-off.
\end{abstract}


\section{Introduction}\label{sec: intro}
With the explosive growth of data and the rising concern of privacy protection, the conventional approach of transmitting and aggregating raw data has become increasingly impractical due to its high bandwidth costs and the significant risks of privacy leakage. Federated learning (FL) \cite{fedavg,mcmahan2017communication,konevcny2016federated,konevcny2016federated_new} emerges as a groundbreaking paradigm, enabling collaboratively model training without sharing private data. However, FL systems face significant security challenges, particularly from scenarios involving \textit{semi-honest} adversaries who follow FL protocols yet analyze the exchanged information, such as model updates, to infer sensitive client information. Among these vulnerabilities, "gradient leakage attacks" pose a severe threat, allowing adversaries to reconstruct private training data with pixel-level accuracy. Several studies, including DLG \cite{zhu2019deepleakagegradients}, Inverting Gradients \cite{geiping2020invertinggradientseasy}, Improved DLG \cite{Zhao2020iDLGID}, and GradInversion \cite{yin2021gradientsimagebatchrecovery}, have demonstrated the feasibility of such attacks. Mitigating this vulnerability is crucial for protecting individual privacy and ensuring the broader adoption of FL in privacy-sensitive industries such as healthcare, finance, and IoT.

Early attempts aiming to thwart privacy attacks include homomorphic encryption (HE) \cite{hardy2017private}, secure multi-party computation (MPC) \cite{SecShare-Adi79,SecShare-Blakley79,bonawitz2017practical}, differential privacy (DP) \cite{abadi2016deep}, and gradient compression (GC) \cite{lin2018deep}. HE and MPC can protect private data without jeopardizing model performance, but they incur heavy computation and communication overhead, especially for deep neural networks. In addition, HE and MPC do not secure clients' private data after decryption for the server aggregation \cite{lam21b}. DP and GC protect data privacy by \textit{distorting} (i.e., adding noise or compressing) shared model updates, which typically leads to significantly deteriorated model performance. To obtain the best of both worlds (i.e., privacy and performance), \cite{wei2021gradient,noble22a,zhu2021fine,shen2022performance} leverage fine-grained DP or regularization to mitigate the impact of noise on model performance.

Early approaches to mitigating privacy attacks include homomorphic encryption (HE) \cite{hardy2017private}, secure multi-party computation (MPC) \cite{SecShare-Adi79,SecShare-Blakley79,bonawitz2017practical}, differential privacy (DP) \cite{abadi2016deep}, and gradient compression (GC) \cite{lin2018deep}. While HE and MPC can safeguard private data without compromising model performance, they impose substantial computational and communication overhead, particularly for deep neural networks. Moreover, HE and MPC do not protect private data after decryption during server-side aggregation \cite{lam21b}. In contrast, DP and GC enhance data privacy by \textit{distorting} shared model updates (e.g., adding noise or applying compression), but this often results in significant degradation of model performance. To achieve a better privacy-utility trade-off, recent works \cite{wei2021gradient,noble22a,zhu2021fine,shen2022performance} leverage fine-grained DP or regularization techniques to minimize the adverse effects of the performance degradation caused by noise. However, existing protection mechanisms struggle to prevent performance degradation, making it challenging to preserve test accuracy while ensuring privacy. 

Geiping et al. \cite{geiping2020invertinggradientseasy} observed that data augmentation during model training increases the difficulty of localizing objects when performing gradient leakage attacks to recover original images. However, these attacks still successfully recover the original data from model updates. This is because the perturbations from augmentations are applied directly to the data, not the gradients, making them vulnerable to reverse-engineering. To address this issue, our approach incorporates the distortion generated by data augmentation directly into the model updates, achieving effects similar to DP. Additionally, this distortion ensures the stability of model training while protecting data privacy.

\begin{figure*}
    \centering
    \includegraphics[width=0.6\linewidth]{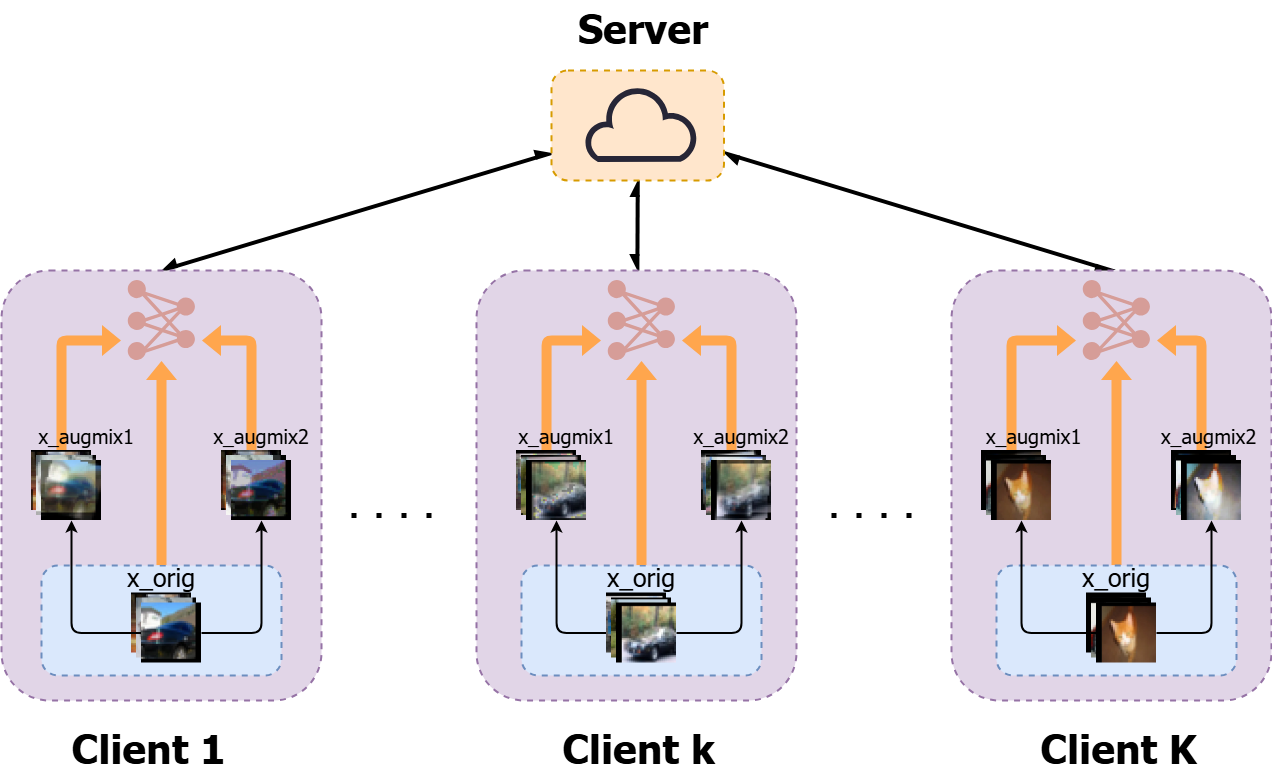}
    \caption{An illustration of training process of Fed-AugMix. Client Updata consists of two parts: (1) In data augmentation part, we generate two augmented data based on original data; (2) In model updating part, we first compute the JS divergence of two augmented data and the original data, then add the divergence to our loss, based on which we update the parameter.}
    \label{fig: overview}
\end{figure*}
In this work, we propose a novel protection framework that integrates a client-level data augmentation algorithm as shown in Figure \ref{fig: overview}. Additionally, we leverage the Jensen-Shannon divergence to introduce controlled distortion into model updates, effectively safeguarding client privacy while maintaining or even enhancing model performance. This approach achieves a favorable balance between privacy and utility. Our contribution are three-fold:

\begin{itemize}
\item  Firstly, we implement the AugMix algorithm (outlined in Algorithm \ref{alg: augmix}) at the client level. This process constructs augmentation chains comprising multiple stochastic operations selected from a predefined set of transformations. The images produced by these chains are then mixed using the MixUp method. To further enhance diversity and consistency, the final output of the augmentation chains is combined with the original image through an additional convex combination. Enable multiple data augmentation methods appear on one image. Employing MixUp ensures that the final mixed image reflects a wide range of transformations.

\item  Secondly, we modify the client-level loss function to incorporate the Jensen-Shannon (JS) divergence between the original and augmented data into the loss function. This modification not only improve the model's robustness but also enhances privacy protection against gradient leakage attacks. By introducing complex noise through the JS divergence, this noise is embedded into the gradients during back-propagation. Due to the inherent complexity of AugMix, this noise becomes challenging to approximate during gradient leakage attacks, significantly hindering data reconstruction. Consequently, the integration of AugMix and JS divergence provides robust privacy preservation against such attacks.

\item  Lastly, we evaluate the proposed algorithm framework on diverse datasets and models to validate its effectiveness. Experimental results reveal that the reconstructed images under our framework's protection are unrecognizable. This demonstrates the robustness of our approach in safeguarding data privacy. Furthermore, our framework consistently achieves a favorable privacy-utility trade-off, and in some cases, it even enhances model performance, highlighting its dual benefits in privacy preservation and model optimization.
\end{itemize}

\section{Related Work}
Relevant prior works consist of 3 parts: study of various data augmentation strategies, gradient attack models, and privacy-utility trade-off in FL.

\noindent\textbf{Data Augmentation.}
Data augmentation is a crucial technique for enhancing model generalization performance. Common methods for image data include random flipping and cropping, which are widely used in practice \cite{resnet}. Mixup offers a distinct approach by combining information from two images through an elementwise convex combination rather than region replacement, which has proven effective for improving model robustness \cite{Zhang2017mixupBE, Tokozume2017BetweenClassLF}. An adaptive mixing policy can further refine Mixup’s effectiveness by mitigating manifold intrusion issues \cite{GuoMixup}. In addition to these manually designed techniques, learned augmentation methods like AutoAugment \cite{autoaugment} optimize a sequence of operations—such as translation, rotation, and shearing—by fine-tuning the probabilities and magnitudes of each operation, ultimately enhancing performance on downstream tasks. In this paper, we implement AugMix \cite{augmix}, a method that enhances model robustness and accuracy on standard benchmark datasets. AugMix achieves this by combining stochastic, diverse augmentations with a Jensen-Shannon Divergence consistency loss and a novel formulation to mix multiple augmented images. 

Furthermore, De Luca et al. \cite{deluca2022mitigatingdataheterogeneityfederated} demonstrate that appropriate data augmentation can mitigate data heterogeneity in FL settings, leading to improved accuracy on unseen clients. Similarly, FedM \cite{fedm-bne}, a data augmentation method based on MixUp-style training, enhances FL performance by enabling data augmentation without requiring the exchange of raw local data among participants.

\noindent\textbf{Gradient Attack Model.}
Despite the privacy-preserving nature of FL, research has shown that shared model updates can still inadvertently leak sensitive information about participants’ private data. Zhu et al. \cite{zhu2019deepleakagegradients} introduced the Deep Leakage from Gradients (DLG) attack, which reconstructs training data by solving an optimization problem on the gradients. Building on this, Geiping et al. \cite{geiping2020invertinggradientseasy} developed the Inverting Gradients (InvGrad) attack, which improves reconstruction quality by utilizing a cosine similarity-based approach. Yin et al. \cite{yin2021gradientsimagebatchrecovery} later proposed the Recursive Gradient Inversion (RGI) attack, which refines data reconstruction by iteratively inverting gradients across multiple rounds. 

In addition to these gradient-based attacks, federated learning systems are vulnerable to other privacy attacks. Notable examples include membership inference attacks, which determine if a specific data point was part of the training set \cite{Nasr_2019}; property inference attacks, which extract general properties of the training data \cite{melis2019}; and model inversion attacks, which aim to approximate sensitive training data directly from model outputs \cite{Fredrikson2015ModelIA}. These findings underscore the importance of developing robust defenses against privacy risks in federated learning.

\noindent\textbf{Privacy-Utility Trade-off.}
Earlier works \cite{reed1973information, yamamoto1983source, sankar2013utility} explored the rate-distortion-equivocation region, quantifying utility by accuracy and privacy by entropy for large data samples. Wang et al. \cite{wang2016relation} assessed privacy leakage through identifiability, differential privacy, and mutual-information frameworks within a unified privacy-distortion model, while Wang et al. \cite{wang2017estimation} proposed a $\chi^2$-based information framework for balancing utility and privacy.

In federated learning, privacy-utility trade-offs are often framed as constrained optimization problems, minimizing utility loss under privacy constraints \cite{zhang2022trading}. Pittaluga et al. \cite{pittaluga2019} used adversarial optimization to train a privacy-preserving encoder within a deep neural network. FL-APB \cite{Liu2024FLAPBBP} combines adversarial training with adaptive privacy protection, dynamically balancing privacy and performance. Zhang et al. \cite{zhang2023theoreticallyprincipledfederatedlearning} propose an algorithmic framework using projected gradient descent to optimize a hyperparameter for near-optimal utility while respecting privacy constraints.

\section{Preliminaries}
In this section, we provide a notation table, formally define the general FL optimization problem, and present the model update framework underpinning our methodology.
\begin{table}[]
    \centering
    \caption{Notations.}
    \begin{tabular}{cc}
    \textbf{Notation} & \textbf{Description} \\
    \hline
    $K$ & Client number \\
    $C$ & Participation rate \\
    $T$ & Communication rounds \\
    $E$ & Local epochs \\
    $w$ & Model parameter \\
    $\mathcal{D}^k$ & Dateset for client $k$ \\
    $x$ & Image \\
    $y$ & Label \\
    $p(y | x)$ & The prediction of image $x$ with model $p(w,x)$\\
    $n$ & Augmentation chain number \\
    $l$ & Augmentation chain length \\
    $s$ & Augmentation severity \\
    $ch_i()$ & $i$th augmentation chain \\
    \hline
    \end{tabular}
    \label{tab: symbol}
\end{table}

The target of FL is to obtain a global model that is collectively trained by clients. It can be formulated as follows:
\begin{equation}
    w^* = \arg\min_{w}\sum_{k = 1}^K \frac{n^{(k)}}{n}\ \mathcal{L}^{(k)}(w)
\end{equation}
where $n^{(k)}$ denotes the size of the dataset $\mathcal{D}^{(k)}$, $n=\sum_{k=1}^K n^{(k)}$, and $\mathcal{L}^{(k)}(w)$ represents the loss of predictions made by the model $w$ on dataset $\mathcal{D}^{(k)}$.

The traditional FL algorithm is FedAvg \cite{fedavg}. The training procedure is described as follows:

\begin{itemize}
    \setlength{\itemsep}{3pt}
    \item Upon receiving the global model $w_{t}$ at communication round $t$, each selected client $k$ locally updates its model parameters over $E$ local iterations, following the rule: $w_{t+1}^{(k)} \leftarrow w_{t}-\eta\cdot\nabla\mathcal{L}^{(k)}(w_{t})$.
    \item After completing the local training, each client transmits its updated model parameters $w_{t+1}^{(k)}$ back to the central server.
    \item Upon receiving the model parameters from all sampled clients, the server aggregates them by computing $w_{t+1} \leftarrow \sum_{k=1}^{K} \frac{n^{(k)}}{n} w_{t+1}^{(k)}$ and subsequently distributes the updated parameters $w_{t+1}$ to all clients.
\end{itemize}

In this context, $w_t$ represents the aggregated model parameters at communication round $t$, and $\eta$ refers to the learning rate. The term $E$ represents the number of local epochs, indicating that each client updates its local model parameters for $E$ iterations before sending the updated parameters back to the server for aggregation.

\section{Method}
We introduce the Fed-AugMix framework in Sec. \ref{sec: framework}, detailing the implementation of AugMix and JS loss in Secs. \ref{sec: augmix} and \ref{sec: js_loss}, respectively. Sec. \ref{sec: loss_scaling} covers our loss scaling technique, and Sec. \ref{sec: trade-off} explains how these methods balance the trade-off between performance and privacy.

\subsection{Framework Overview}\label{sec: framework}
To tackle the trade-off between performance and privacy protection, as mentioned in Sec. \ref{sec: intro}, we present a data augmentation framework for FL: Fed-AugMix, aimed at enhancing model performance and robustness while providing privacy protection. The learning procedure is illustrated in Figure \ref{fig: overview}, and the full training process of Fed-AugMix is shown in Algorithm \ref{alg: Fed-AugMix}. Specifically, we adopt the AugMix algorithm \cite{augmix} on clients, as detailed in Algorithm \ref{alg: augmix}.

At the server level, we aggregate the weights of selected client models. At the client level, we stochastically augment the training data two times and make some adaptations to the loss function. Initially, stochastic augmentation operations are applied, layered, and combined to generate a diverse set of augmented images. Additionally, we use the MixUp technique \cite{mixup} to further process both the augmented images and the original image. Besides, we incorporate the Jensen-Shannon (JS) divergence into the loss function.

\begin{algorithm}[t]
\caption{Fed-AugMix.}
\textbf{Input:} $K$: client number; $C$: the fraction of active client in each round; $T$: communication rounds; $E$: local epochs; $\eta$: learning rate; $w$: model parameters; $f(w,x)$: model's output of image $x$; $\mathcal{L}_c$: classification loss; $\mathcal{D}$: datasets for clients, $\mathcal{D} = \{\mathcal{D}^{1},\cdots, \mathcal{D}^{(K)}\}$\par
\textbf{Server executes:}
\begin{algorithmic}[1]
    \STATE Initialize $w_0$
    \FOR{each round $t = 1, 2, \dots, T$ }
        \STATE $m \gets \max(C \cdot K, 1)$
        \STATE $S_t \gets$ (random set of $m$ clients)
        \FOR{each client $k \in S_t$ \textbf{in parallel}}
            \STATE $w_{t+1}^{k} \gets$ \textsc{ClientUpdate}$(k, w_t)$
        \ENDFOR
        \STATE $w_{t+1} \gets \textsc{GlobalUpdate}(w_{t+1}^{1}, \cdots, w_{t+1}^{K})$
    \ENDFOR
\end{algorithmic}

\vspace{1em}

\textbf{ClientUpdate}$(k, w)$: \AlgComment{Run on client $k$}\par
\begin{algorithmic}[1]
    \STATE Initialize $\mathcal{L}^{(k)}(w)$
    \FOR{each local epoch $i$ from $1$ to $E$}
        \FOR{( image $x_\text{orig}$ and label $y$ ) $\in \mathcal{D}^\text{(k)}$}
            \STATE $x_\text{augmix1} = \text{AugMix}(x_\text{orig})$
	    \STATE $x_\text{augmix2} = \text{AugMix}(x_\text{orig})$ 
	    \AlgComment{$x_\text{augmix1}\ne x_\text{augmix2}$}
            \STATE $p_\text{orig} = p(y \mid x_\text{orig}) = f(w, x_\text{orig})$
            \STATE $p_\text{augmix1} = p(y \mid x_\text{augmix1}) = f(w, x_\text{augmix1})$
            \STATE $p_\text{augmix2} =p(y \mid x_\text{augmix2}) = f(w, x_\text{augmix2})$
            \STATE $\mathcal{L}^{(k)}(w) = \mathcal{L}_c(p_\text{orig}, y)+ \lambda D_\text{JS}(p_\text{orig}; p_\text{augmix1}; p_\text{augmix2})$
        \STATE $w \gets w - \eta \nabla \mathcal{L}^{(k)}(w)$         
        \ENDFOR
    \ENDFOR
    \RETURN{$w$ to server}
\end{algorithmic}
\label{alg: Fed-AugMix}
\end{algorithm}
\begin{algorithm}[]%
    \textbf{Input:} $x_\text{orig}$: original image; $n$: augmentation chain number; $\mathcal{O}$: operation set, $\mathcal{O} = \{\text{rotate}, \text{shear}, \ldots, \text{posterize}\}$; $s$: augmentation severity.\par
    \begin{algorithmic}[1]
		\STATE Fill $x_\text{aug}$ with zeros
		\STATE Sample mixing weights $(b_1, \ldots, b_n) \sim \text{Dirichlet}(\alpha,\ldots,\alpha)$
		\FOR{$i = 1, \ldots, n$}
        \STATE Randomly choose chain length $l$ from 1 to 3.
        \STATE Sample operations $\text{op}_1, \cdots, \text{op}_l \sim \mathcal{O}$
        \STATE Compose operation chain $ch_i$ with operations of length $l$, $ch_i = \text{op}_{1\cdots l} = \text{op}_l \circ \cdots \circ \text{op}_1$
        \STATE $x_\text{aug} \mathrel{+}= b_i \cdot ch_i(x_\text{orig},\ s)$ \AlgComment{Addition is elementwise}
		\ENDFOR
		\STATE Sample weight $m \sim \text{Beta}(\alpha, \alpha)$
		\STATE Interpolate with rule $x_\text{augmix} = m x_\text{orig} + (1 - m) x_\text{aug}$
		\RETURN{$\,x_\text{augmix}$}
		\end{algorithmic}
	\caption{AugMix.}
	\label{alg: augmix}
\end{algorithm}

\subsection{AugMix}\label{sec: augmix}
\begin{figure*}
    \centering
    \includegraphics[width=0.8\linewidth]{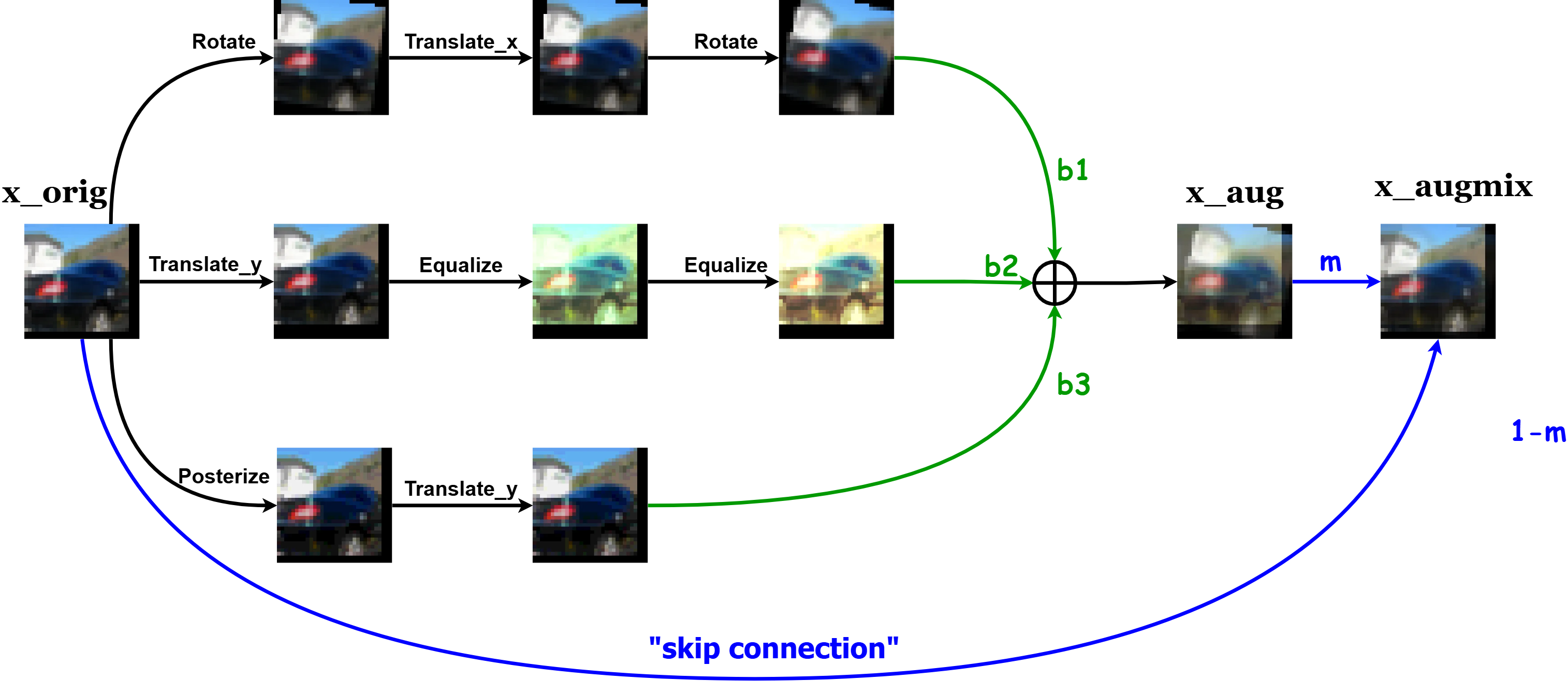}
    \caption{An example of AugMix. First generate x\_aug using three stochastic augmentation chains. Then employ "skip connection" to MixUp the augmented image and the original image.}
    \label{fig: augmix}
\end{figure*}

In this section, we provide a comprehensive explanation of how the AugMix algorithm utilizes two techniques: data augmentation\cite{data_augmentation} and MixUp. An example of the procedure of AugMix is illustrated in Figure \ref{fig: augmix}.

First, we employ augmentation chains composed of multiple stochastic operations selected from a predefined set of augmentations. Specifically, we sample operations from AutoAugment \cite{autoaugment} to construct an operation chain. We generate $n$ independent operation chains to produce $n$ different augmented images, where $n=3$ by default. For example, in the first row of Fig. \ref{fig: augmix}, the first augmentation chain consists of $Rotate$, $Translate\_x$ and another $Rotate$, each with varying augmentation level. These two $Rotate$ operations can vary in rotation angle ranging from $2^\circ$ to $-15^\circ$, where the augmentation level $AugLevel$ (rotation angle in this case) is determined based on sample level $SampLevel$ and a predefined max value $MaxVal$ that vary in operation:
\begin{align}
    \text{AugLevel} = \frac{\text{SampLevel}}{10}\cdot \text{MaxVal}\ \text{,}
\end{align}
where $SampLevel \sim \mathcal{U}(0.1,\ s)$ and hyperparameter $s$ denotes augmentation severity.

After the augmented images are generated through operation chains, they are combined using a convex mixing process. Instead of alpha compositing, we use elementwise convex combinations for simplicity. First, the $n$ augmented images $ch_i(x_{\text{orig}},\ s)$ are mixed to produce a final output $x_{\text{aug}}$ based on
\begin{align}
    x_{\text{aug}} = \sum_i^n b_i \cdot ch_i(x_{\text{orig}},\ s)\ \text{,}
\end{align}
where $(b_1, \ldots, b_n) \sim \text{Dirichlet}( \alpha,\ldots,\alpha)$. Moreover, we combine the final output of the augmentation chains $x_\text{aug}$ with the original image $x_\text{orig}$ through a second convex combination, known as "skip connection":
\begin{align}
    x_\text{augmix} = m x_\text{orig} + (1 - m) x_\text{aug}\ \text{,}
\end{align}
where $m \sim \text{Beta}(\alpha, \alpha)$. Employing MixUp ensures that the final mixed image reflects a wide range of transformations, incorporating several sources of perturbation: the choice of operations, the severity of these operations, the lengths of the augmentation chains and the mixing weights.

\subsection{JS (Jensen-Shannon) Loss}\label{sec: js_loss}
To ensure model consistency across different augmentations of the same image, we modify the client-level loss function. Since the semantic content of an image is preserved under AugMix, the model is expected to learn the core representations while ignoring the perturbations introduced during the AugMix process. Thus, we aim for similar predicted probabilities for the original image $x_\text{orig}$ and its two stochastically augmented versions $x_\text{augmix1}$ and $x_\text{augmix2}$. To achieve this, we introduce the JS divergence into the loss function. Given the model's predictions of posterior probability distribution $p_\text{orig} = p(y \mid x_\text{orig})$, $p_\text{augmix1} = p(y \mid x_\text{augmix1})$ and $p_\text{augmix2} = p(y \mid x_\text{augmix2})$, the original loss function $\mathcal{L}_c$ is modified to:
\begin{equation}
\mathcal{L}_c(p_\text{orig},\ y) + \lambda \, D_\text{JS}(p_\text{orig};\ p_\text{augmix1};\ p_\text{augmix2}).
\end{equation}

To compute the JS divergence $D_\text{JS}$, we first calculate the average distribution \( q \) as \( q = (p_\text{orig} + p_\text{augmix1} + p_\text{augmix2}) / 3 \). Next, we compute the KL divergence between \( q \) and each of the three distributions: \( p_\text{orig} \), \( p_\text{augmix1} \) and \( p_\text{augmix2} \). Finally, we take the average of these KL divergences to obtain the JS divergence:
\begin{align}
D_\text{JS}(p_\text{orig};\ p_\text{augmix1};\ p_\text{augmix2}) = \frac{1}{3}\Bigl(\ D_\text{KL}(p_\text{orig}\ \| \ q&) \nonumber \\
 \quad + D_\text{KL}(p_\text{augmix1}\ \| \ q&) \nonumber \\
 \quad + D_\text{KL}(p_\text{augmix2}\ \| \ q&)\ \Bigr).
\end{align}

Unlike KL divergence, JS divergence is bounded above by the logarithm of the number of classes. We select two augmented images for sampling because computing \( D_\text{JS}(p_\text{orig}; p_\text{augmix1}) \) alone underperforms compared to our approach. Adding more distributions, such as \( D_\text{JS}(p_\text{orig}; p_\text{augmix1}; p_\text{augmix2}; p_\text{augmix3}) \), increases computational cost with only marginal performance gains. The Jensen-Shannon Consistency Loss thus encourages model stability, consistency, and resilience to input variations \cite{Bachman, Zheng_2016, alp}.

\subsection{Balancing Performance and Privacy}\label{sec: trade-off}
The core innovation of our framework is the incorporation of JS divergence in the loss function, which evaluates the local model's ability to maintain consistent probability distributions across stochastic augmentations of the same data. we provide an intuitive explanation of our framework through two key questions: \textbf{\underline{Q1:}} How does Fed-AugMix improve model's performance? \textbf{\underline{Q2:}} How does Fed-AugMix offer privacy protection against gradient leakage attack (GLA), such as InvGrad attack?

To enhance model performance, AugMix introduces stochastic data corruption through random augmentation operations, varying severities, chain lengths, and mixing weights. Rather than training directly on augmented data, we incorporate JS divergence to measure prediction disparities between the original and two augmented versions of the same data. This approach enforces consistency in local model predictions across augmentations, encouraging models to learn effective representations while disregarding perturbations from AugMix. Consequently, this improves model generalization in FL scenarios and enhances robustness against various corruptions.

To explain Fed-AugMix’s privacy protection mechanism, it’s essential to first understand GLA. GLA works by exploiting the gradient-sharing process in collaborative training. While normal participants compute gradients using their private training data, a malicious attacker initialize “dummy inputs” (random noise) and corresponding labels, iteratively updating them to minimize the gradient distance to the shared gradients. By optimizing these dummy gradients to fit the leaked gradient, the attacker can effectively reconstruct the input data. 

A effective method to mitigate privacy leakage risk posed by GLA is Differential Privacy (DP), which adds random noise on gradients. This noise reduces the precision of gradient information, making it difficult for attackers to reconstruct original data, while simultaneously limiting the influence of individual data points on the gradient, thereby preserving privacy. Our framework, Fed-AugMix, shares certain similarities with DP.

 Since GLA can reconstruct original data from gradients, simply data augmentation alone remains vulnerable to such recovery. Although AugMix introduces perturbations, an attacker could potentially reverse-engineer these modifications to reconstruct the augmented data \cite{geiping2020invertinggradientseasy}. As standard augmentations like AugMix do not inherently render data indistinguishable, augmented data remains at risk of recovery, potentially leading to privacy leakage of the original data.

In addition to AugMix, our framework introduces a JS loss that integrates JS divergence into the loss function, effectively enhancing privacy protection against GLA. Since AugMix generates stochastic augmentations, the JS divergence between the original and augmented data can be viewed as complex noise. Through back-propagation, this noise is incorporated into each gradient. Due to the complexity of AugMix, this noise is difficult to aprroximate during GLA, making data reconstruction challenging. Thus, the combination of AugMix and JS divergence offers robust privacy preservation against attacks.

\subsection{Scaling the Loss}\label{sec: loss_scaling}
In the InvGrad attack experiment discussed in Sec. \ref{sec: privacy}, we find that our framework does not fully protect privacy, as a small portion of the image can still be recovered, revealing information about the original data. To identify factors limiting privacy protection, we analyze the algorithm and observe that, when attacking an untrained model, the classification loss \(\mathcal{L}_c\) is approximately equal to the JS divergence scaled by \( 10^5 \). Given the negligible magnitude of the JS divergence, the noise added to gradients through back-propagation remains minimal, making data recovery feasible. 

As we increase \( \lambda \) (the coefficient of JS divergence) to a large value, such as \( 10^5 \), the model's performance degrades significantly. We attribute this to the modified loss function, which causes the model to prioritize optimizing the JS divergence over the \(\mathcal{L}_c\) during training. Specifically, during model training, while the \(\mathcal{L}_c\) decreases, \( \lambda D_{\text{JS}} \) remains high, leading subsequent training to primarily focus on JS divergence. Consequently, this disrupts convergence and hampers model performance.

To address this dilemma, we propose a phased approach to adjusting \( \lambda \). Here, \(\mathcal{L}_c\) denotes the classification loss, \(D_\text{JS}\) represents the Jensen-Shannon divergence, and \( Scale \) is a hyperparameter controlling the sensitivity of the condition. When \(\mathcal{L}_c\) surpasses the scaled divergence, \(\lambda\) is set to a large constant value \( LargeVal \); otherwise, \(\lambda\) retains its previous value. In the early training stages, when JS divergence is low, we set \( \lambda \) to a higher value $LargeVal$ to ensure sufficient noise is added to the gradients for privacy protection. In later stages, as lower loss provides less gradients information for attackers to exploit, we reduce \( \lambda \) to prioritize optimizing the \(\mathcal{L}_c\), thereby maintaining model performance. This approach effectively balances performance with privacy requirements.
\section{Experiments}
In this section, we evaluate the Fed-AugMix framework's ability to balance performance and privacy. First, we analyze the privacy protection achieved under varying augmentation severities and compare it with the baseline of no protection. Next, we present results on three datasets, demonstrating that Fed-AugMix delivers a superior privacy-utility trade-off, effectively safeguarding privacy while preserving model performance. Finally, we show empirical evidence of improved test accuracy and faster convergence with our framework in certain scenarios.

    
    
    
\begin{table}[]
\centering
\caption{The average of defense effect metrics on MNIST, CIFAR10 and CIFAR100 (measured by MSE, SSIM and PSNR). Here, an upward arrow (\(\uparrow\)) indicates that higher values correspond to better protection, while a downward arrow (\(\downarrow\)) signifies that lower values are preferred for improved defense outcomes.}
\begin{tabular}{@{}c|c|c|c|c@{}}
\multicolumn{5}{c}{\textbf{MNIST}}\\
\toprule
Stage & Protection & MSE ($\uparrow$) & SSIM ($\downarrow$) & PSNR ($\downarrow$)\\ \midrule
\multirow{6}{*}{\textsc{untrained}} 
    & none      & 1.387 & 10.23\% &  9.204\\ 
    & $s=2$      & 2.547 & 2.54\% &  6.519\\ 
    & $s=4$      & 2.749 & 2.64\% &  6.186\\
    & $s=6$      & 2.811 & 2.08\% & 6.126 \\ 
    & $s=8$      & 2.838 & 2.71\% & 6.142 \\ 
    & $s=10$      & \textbf{2.909} & \textbf{1.94}\% & \textbf{5.992} \\ \midrule
\multirow{6}{*}{\textsc{convergent}} 
    & none      & 2.092 & 6.65\% &  7.731\\ 
    & $s=2$      & 2.260 & 4.02\% &  7.237\\     
    & $s=4$      & 2.292 & 3.36\% & 7.142 \\    
    & $s=6$      & 2.377 & \textbf{2.56}\% & 6.919 \\     
    & $s=8$      & 2.382 & 2.71\% & 6.965 \\     
    & $s=10$      & \textbf{2.414} & 3.36\% & \textbf{6.883} \\
\midrule
\addlinespace[2em] 
\multicolumn{5}{c}{\textbf{CIFAR10}}\\
\midrule
Stage & Protection & MSE ($\uparrow$) & SSIM ($\downarrow$) & PSNR ($\downarrow$)\\ \midrule
\multirow{6}{*}{\textsc{untrained}} 
    & none      & 3.445 & 1.08\% &  8.744\\ 
    & $s=2$      & 3.725 & 0.73\% &  8.339\\
    & $s=4$      & 3.786 & 0.66\% &  8.266\\ 
    & $s=6$      & 3.828 & 0.68\% &  8.222\\
    & $s=8$      & 3.835 & 0.62\% & \textbf{8.208}\\ 
    & $s=10$      & \textbf{3.836} & \textbf{0.59\%} & 8.212 \\ \midrule
\multirow{6}{*}{\textsc{convergent}} 
    & none      & 3.629 & 1.94\% & 8.584\\ 
    & $s=2$      & 3.751 & 1.56\% & 8.387\\     
    & $s=4$      & 3.707 & 1.68\% & 8.481 \\    
    & $s=6$      & 3.760 & 1.42\% & 8.411 \\     
    & $s=8$      & 3.837 & 1.33\% & 8.328 \\     
    & $s=10$      & 4.323 & 1.01\% & 7.785 \\
\midrule
\addlinespace[2em] 
\multicolumn{5}{c}{\textbf{CIFAR100}}\\
\midrule
Stage & Protection & MSE ($\uparrow$) & SSIM ($\downarrow$) & PSNR ($\downarrow$)\\ \midrule
\multirow{6}{*}{\textsc{untrained}} 
    & none      & 3.462 & 0.95\% &  8.722\\ 
    & $s=2$      & 3.817 & 0.76\% &  8.272\\
    & $s=4$      & 3.857 & 0.75\% &  8.212\\ 
    & $s=6$      & 3.926 & \textbf{0.64\%} &  8.141\\
    & $s=8$      & 3.968 & 0.72\% & 8.085 \\ 
    & $s=10$      & \textbf{3.991} & 0.66\% & \textbf{8.065} \\ \midrule
\multirow{6}{*}{\textsc{convergent}} 
    & none      & 3.096 & 3.64\% &  9.345\\ 
    & $s=2$      & \textbf{3.275} & 2.87\% &  \textbf{9.063}\\     
    & $s=4$      & 3.195 & 2.57\% & 9.147 \\    
    & $s=6$      & 3.104 & 2.67\% & 9.267 \\     
    & $s=8$      & 3.131 & 2.18\% & 9.223 \\     
    & $s=10$      & 3.011 & \textbf{2.08\%} & 9.393 \\
\bottomrule
\end{tabular}
\label{tab: protection-metrics}
\end{table}
\subsection{Experimental Setup}
\noindent\textbf{Dataset Partitioning.} We conducted experiments on three datasets: MNIST \cite{lecun2010mnist}, CIFAR10 \cite{cifar} and CIFAR100 \cite{cifar}. We set the number of clients to $K=100$, with $10\%$ of clients ($C=0.1$) participating in each communication round, meaning 10 clients were active per round. To simulate a non-IID data distribution, we used Dirichlet sampling with a parameter of $\alpha=0.1$ to partition the data across clients. Each client's data was further divided into training and test sets with a test split ratio of 0.25.

\noindent\textbf{Model Architectures.}
To address the varying image sizes, dataset scales, and task complexities, we employed two distinct neural network architectures. For the MNIST and FMNIST datasets, we utilized LeNet-5 \cite{lenet}, a 5-layer neural network suited for simpler datasets. For the CIFAR-10 and CIFAR-100 datasets, we adopted ResNet-50 \cite{resnet}, a deeper architecture designed for more complex classification tasks. This selection ensures that the model capacity aligns with the complexity of each dataset.

\noindent\textbf{Attack Setting.}
We adopt the InvGrad method \cite{geiping2020invertinggradientseasy} as the attack strategy. Consistent with common attack assumptions in federated learning, the attacker is presumed to have access to weight updates. Additionally, we assume the attacker has knowledge of the true labels to simplify the attack. Experiments are conducted on MNIST, CIFAR-10, and CIFAR-100, as these datasets are relatively easier to reconstruct. For each target batch, 2500 attack iterations are performed using the Adam optimizer with a learning rate of 0.1 and a total variation coefficient of $1 \times 10^{-6}$. Given the limitations of gradient leakage attacks in large-batch attacks \cite{zhu2019deepleakagegradients}, we use a small batch size of 4 and set the local training epoch to 5 to increase the vulnerability of the training process. For each of the 100 clients, the attack is performed on a specific batch at two distinct training stages: \textsc{untrained} and \textsc{convergent}.

\noindent\textbf{Privacy Setting.}
We vary the augmentation severity across a range of values to explore the trade-off between performance and privacy. In loss scaling procedure, we set $Scale$ to $5 \times 10^4$, $LargeVal$ to $5 \times 10^3$, and $\lambda$ to 50, which effectively mitigate privacy leakage, particularly during the early stages of training. The defense effectiveness is evaluated using metrics such as Mean Squared Error (MSE), Structural Similarity Index (SSIM), and Peak Signal-to-Noise Ratio (PSNR).

\subsection{Results and Analysis}\label{sec: privacy}
\begin{figure*}[]
	\centering
	\begin{minipage}{0.49\linewidth}
		\centering
		\includegraphics[width=1\linewidth]{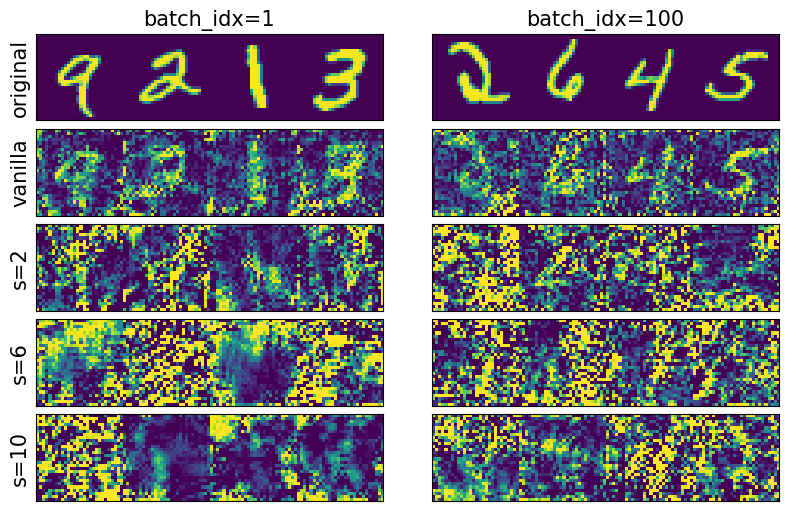}
	\end{minipage}
	\begin{minipage}{0.49\linewidth}
		\centering
		\includegraphics[width=1\linewidth]{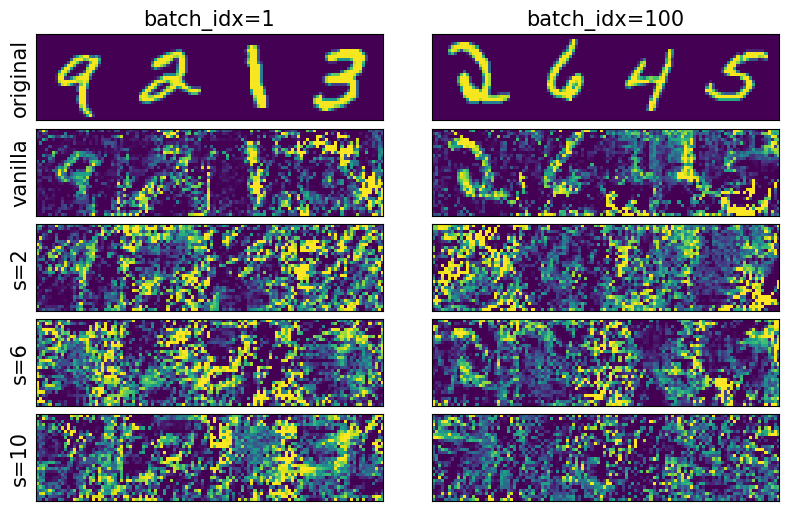}
	\end{minipage}
    \caption{Visualization of InvGrad attack results under varying privacy protection severity levels (s=0, 2, 6, 10). The second row shows reconstructions without protection, while the following rows display reconstructions with Fed-AugMix at different augmentation severities.}
    \label{fig: protection_over_severity}
\end{figure*}
\noindent\textbf{Protecting Privacy.}
We evaluated privacy protection on the MNIST dataset under varying stages and levels of protection severity using metrics such as Mean Squared Error (MSE), Structural Similarity Index Measure (SSIM), and Peak Signal-to-Noise Ratio (PSNR). Figure \ref{fig: protection_over_severity} presents the results of our privacy protection experiments. Without protection, the original images expose a notable amount of private information when subjected to the InvGrad attack. However, integrating Fed-AugMix and Loss Scaling significantly reduces privacy leakage, rendering the reconstructed images unrecognizable.

Across different stages and levels of protection severity, as summarized in Table \ref{tab: protection-metrics}, our method consistently demonstrates a high level of effectiveness. Notably, the SSIM values remain below 5\%, indicating minimal similarity between the reconstructed and original images. Furthermore, as the severity of augmentations increases, the MSE also rises, reflecting an enhanced degree of privacy preservation over protection severity.
\begin{figure}[H]
    \centering
    \includegraphics[width=\linewidth]{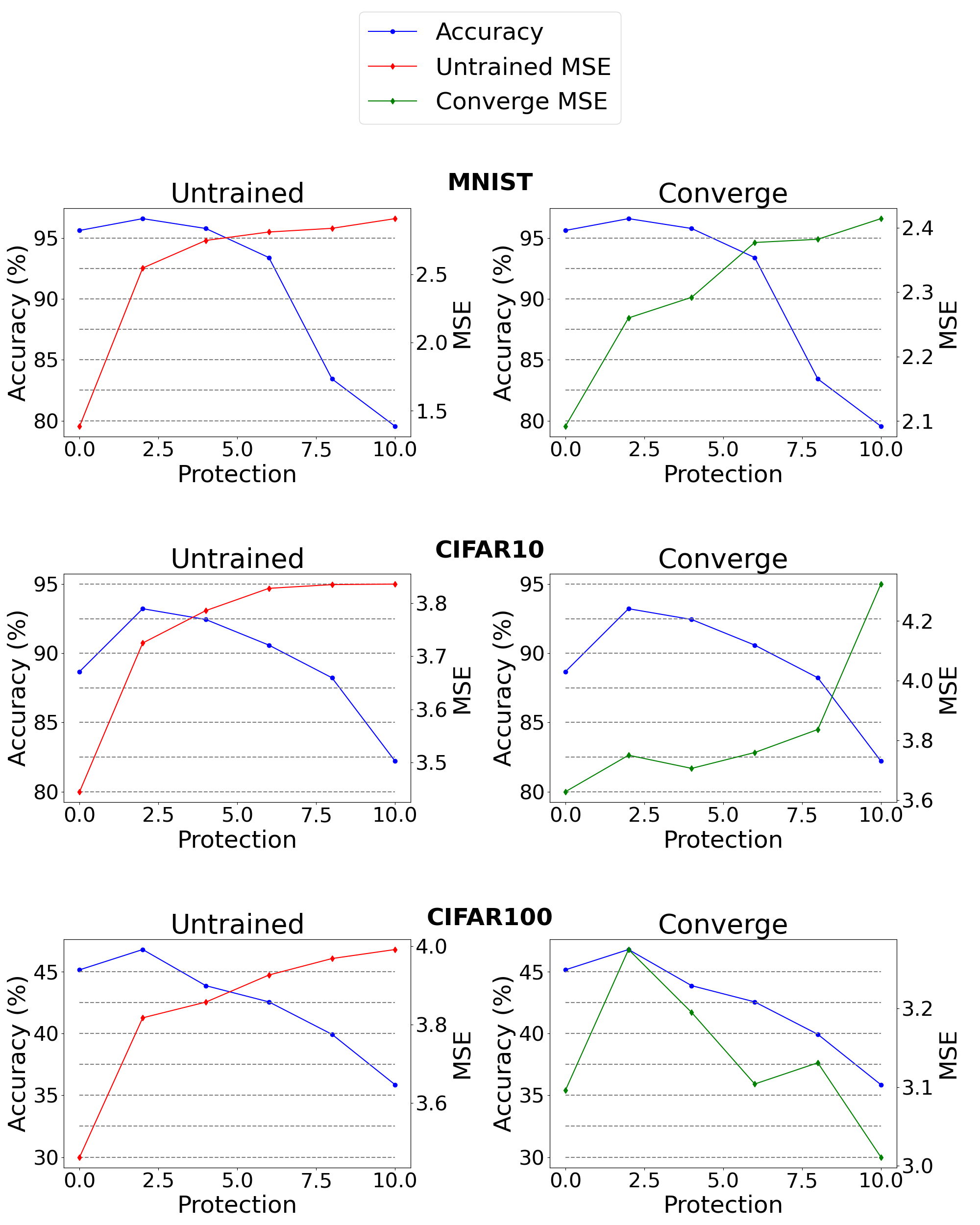}
    \caption{Relationship between test accuracy and MSE of reconstructed images under varying protection severity. Lower severity reduces MSE while improving accuracy compared to no protection. However, as augmentation severity increases, accuracy decreases for both untrained and converged models.}
    \label{fig: acc_mse}
\end{figure}

\begin{figure*}
    \centering
    \includegraphics[width=1\linewidth]{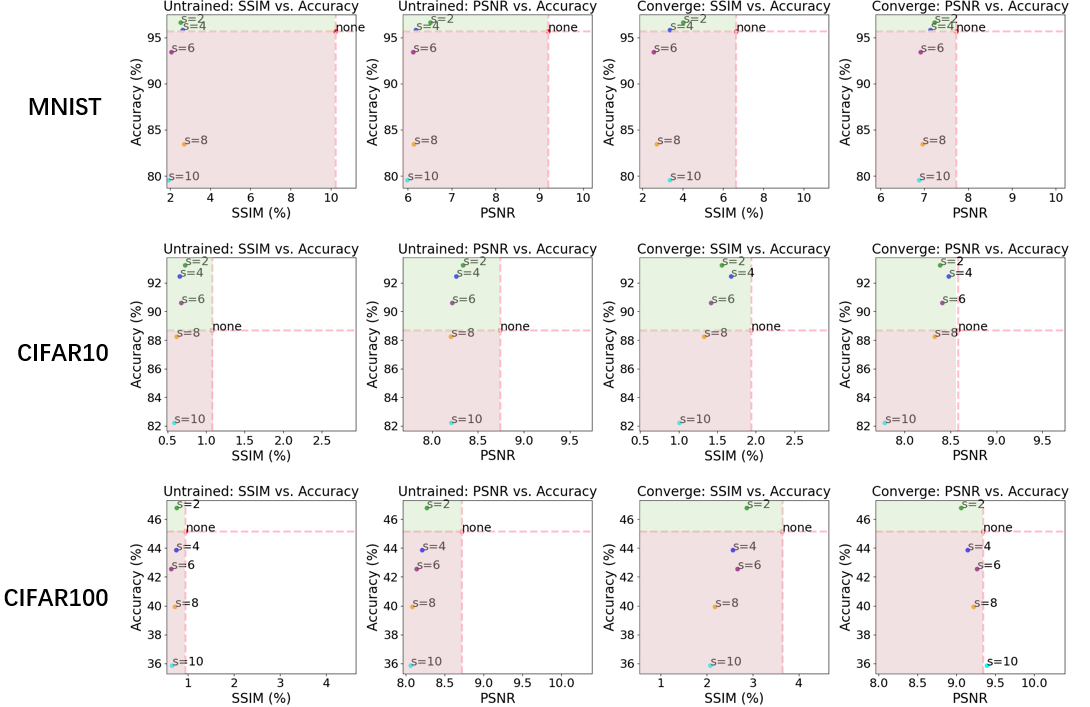}
    \caption{An illustration of the relationship between accuracy, SSIM, and PSNR for untrained and converged models across different datasets. The green region indicates improved accuracy alongside better privacy protection, while the red region reflects enhanced privacy protection at the cost of performance degradation. Most privacy protection mechanisms of FL fall within the red region.}
    \label{fig: tradeoff}
\end{figure*}
\noindent\textbf{Privacy-Utility Trade-off.}
We conduct experiments to evaluate test accuracy under varying levels of augmentation severity and analyze the privacy-utility trade-off of Fed-AugMix using metrics such as MSE, SSIM, and PSNR. As shown in Figure \ref{fig: acc_mse}, test accuracy declines with increasing augmentation severity, while the MSE between original and reconstructed images rises. When employing a converged model for image recovery, raising the augmentation severity from 0 to 10 increases the MSE from 2.092 to 2.414. Similarly, using an untrained model, the MSE increases from 1.387 to 2.909.

The experimental results demonstrate the framework's ability to balance privacy and accuracy, highlighting that the trade-off between privacy and utility can be controlled by adjusting the augmentation severity \( s \). Specifically, applying augmentations with lower severity initially improves model performance. Additionally, the distortions introduced to the images, which propagate back to the gradients, help protect privacy. Figure \ref{fig: tradeoff} illustrates the effectiveness of our framework in achieving this balance by showing SSIM and PSNR versus accuracy across different augmentation severity levels. To enhance clarity, the \textsc{untrained} and \textsc{converged} states are represented as separate lines. In the figure, points in the upper-left corner indicate a favorable trade-off between privacy and utility. Notably, augmentation severities \( s = 2 \), \( s = 4 \), and \( s = 6 \) are located in this region, demonstrating their effectiveness in balancing privacy and utility.
\begin{figure}[H]
    \centering
    \includegraphics[width=1\linewidth]{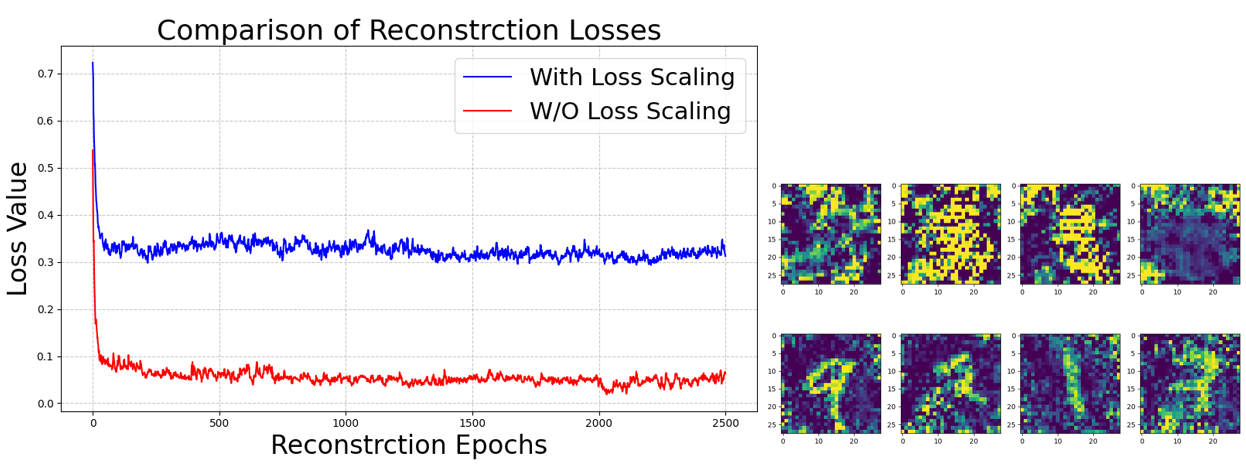}
    \caption{An example demonstrating the effectiveness of Loss Scaling: the reconstruction loss is higher with Loss Scaling than without it, reducing the likelihood of privacy leakage.}
    \label{fig: loss_scale}
\end{figure}
\noindent\textbf{Loss Scaling.}
To address privacy leakage during the early stages of training, we propose Loss Scaling, which increases the JS divergence between different augmentations of the same image. To assess its effectiveness, we conducted two InvGrad attacks on an untrained model over 2500 iterations using the same batch: one with Loss Scaling and one without. As illustrated in Figure \ref{fig: loss_scale}, without Loss Scaling, the attacker achieves a lower reconstruction loss, nearly recovering the original images. In contrast, with Loss Scaling, the reconstructed images become indistinguishable. These results indicate that Fed-AugMix is vulnerable to privacy leakage during early training stages. However, the inclusion of Loss Scaling significantly reduces reconstruction loss, effectively mitigates privacy leakage.

\begin{figure*}
    \centering
    \includegraphics[width=1\linewidth]{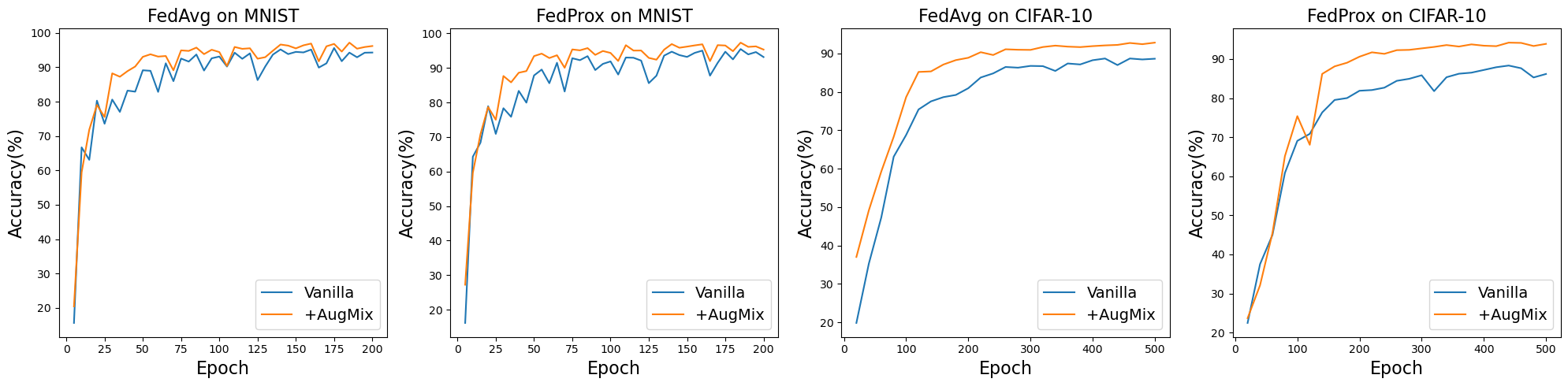}
    \caption{These charts illustrate the test accuracy of two federated learning methods, FedAvg and FedProx, during the training process. The left charts show the test accuracy on the MNIST dataset, comparing the results with and without the use of Fed-AugMix. The right charts display the same comparison on the CIFAR-10 dataset.}
    \label{fig: fedprox}
\end{figure*}

\noindent\textbf{Compatibility and Effectiveness.}
We evaluate the performance of different FL algorithms on multiple datasets, both with and without our proposed Fed-AugMix framework. Table \ref{tab: compatability} highlights the compatibility and effectiveness of our framework across different datasets, including MNIST, CIFAR10 and CIFAR100. As shown in Table \ref{fig: fedprox}, integrating Fed-AugMix with each FL method consistently improves test accuracy across all datasets. For example, when applying FedProx to CIFAR10, the vanilla FedProx achieves a test accuracy of $88.33\%$, while FedProx combined with Fed-AugMix improves to $94.19\%$, representing a $5.86\%$ increase. These results indicate that our framework is highly compatible with a range of FL algorithms and demonstrates significant improvements in performance, making it a versatile solution for diverse applications.

\section{Conclusion and Discussion}
This paper investigates the trade-off between privacy and utility in FL scenarios. To address this challenge, we propose a novel framework, Fed-AugMix, which applies the AugMix algorithm at the client level. This approach enhances model performance, robustness, and generalization while introducing perturbations to the original data that propagate to the gradients, thereby preserving client privacy. Furthermore, we incorporate Loss Scaling to ensure consistent privacy preservation throughout the entire training process.

\noindent\textbf{Limitations.} 
Data augmentation introduces only limited distortions to the images, resulting in relatively minimal noise being added to the gradients. However, differential privacy can overcome this limitation by injecting noise with predefined parameters \(\mu\) and \(\sigma\), allowing precise control over the level of privacy protection. Another limitation is the computational overhead of using AugMix, which typically increases the training time due to its data augmentation operations.

\noindent\textbf{Future Work.}
Future work should focus on providing deeper interpretations and conducting rigorous theoretical analyses of integrating data augmentation into FL frameworks. Additionally, a systematic evaluation of various augmentation strategies is needed to assess their impact on both model performance and privacy. Building on these insights, future research could aim to design more controllable and effective data augmentation methods tailored to FL scenarios.


\bibliographystyle{IEEEtran}
\bibliography{main}

\newpage

\onecolumn
\appendix
\section{Additional Experiment Results}
We evaluated InvGrad attacks on CIFAR-10 and CIFAR-100 datasets using ConvNet, an 8-layer CNN because of the susceptibility to gradient inversion attacks.

\begin{figure}[htbp]
	\centering
	\begin{minipage}{\linewidth}
		\centering
		\includegraphics[width=\linewidth]{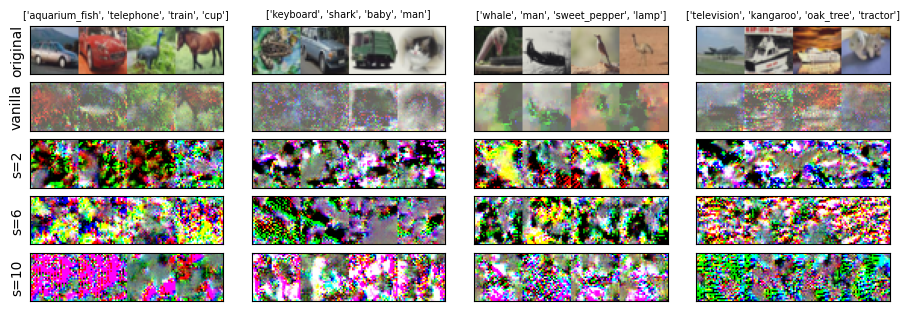}
		\caption{CIFAR10}
		\label{fig: cifar10}
	\end{minipage}

	\qquad
	\begin{minipage}{\linewidth}
		\centering
		\includegraphics[width=\linewidth]{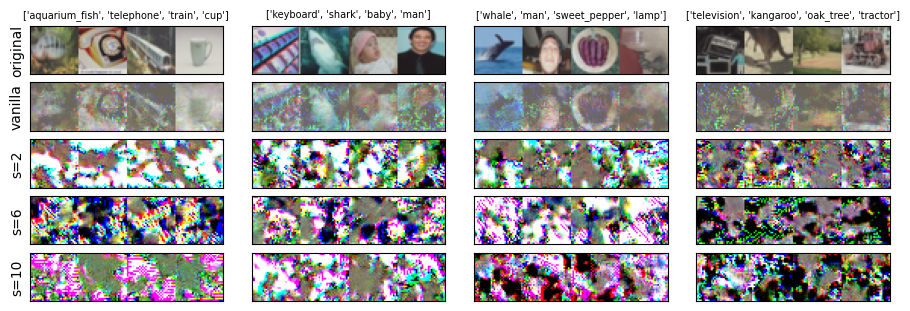}
		\caption{CIFAR100}
		\label{fig: cifar100}
	\end{minipage}

\end{figure}

As Figure \ref{fig: cifar10} and Figure \ref{fig: cifar100} show, without the protection of Fed-AugMix, the reconstructed images reveal significant privacy leakage, as the original content can be effectively recovered. In contrast, applying Fed-AugMix with varying augmentation severities substantially mitigates this risk, effectively preventing the recovery of private image content and enhancing privacy protection.

We evaluated the test accuracy of Fed-AugMix on MNIST, CIFAR-10 and CIFAR-100 datasets. We employed lenet-5 for MNIST 10-class image classification, and ResNet50 for CIFAR10 and CIFAR100.

As Table \ref{tab: compatability} demonstrates, with the introduction of data augmentation and JS divergence in Fed-AugMix, the model performance increase in all 3 datasets. Our framework can employ other FL algorithms as backbone, in order to fit a specific task and function.

\begin{table}[htbp]
\caption{The test accuracy of different FL methods, with and without Fed-AugMix, on the MNIST, CIFAR10, and CIFAR100 datasets with $\rm C=10\%$. We have bolded the highest test accuracy. The numbers within parentheses represent the improvement of accuracy, and $\star$ on the cell means in that case, the model will collapse.}
\label{tab: compatability}
\vskip -0.2in
\begin{center}
\begin{adjustbox}{width=0.8\textwidth}
\begin{tabular}{c|cc|cc|cc} 
\toprule
 \multirow{2}{*}{Dataset}&\multicolumn{2}{c|}{\multirow{2}{*}{MNIST}}                & \multicolumn{2}{c|}{\multirow{2}{*}{CIFAR10}}             & \multicolumn{2}{c}{\multirow{2}{*}{CIFAR100}}                \\
 &&&&&&\\
 \midrule
 Method & vanilla(\%)         & Fed-AugMix(\%)       & vanilla(\%)   & Fed-AugMix(\%)  & vanilla(\%)      & Fed-AugMix(\%)        \\ 
\midrule
 FedAvg   & 95.63           & 96.60    & 88.68          & 93.22      & 45.13              & 46.77     \\
 FedProx  & 95.88           & 97.34    & 88.33     & 94.19 & 45.24          & 47.83         \\
\bottomrule
\end{tabular}
\end{adjustbox}
\end{center}
\vskip -0.2in
\end{table}

\end{document}